\documentclass[aps,prd,amsmath,amssymb,superscriptaddress,twocolumn,nofootinbib,showpacs,preprintnumbers,notitlepage]{revtex4-1}

\usepackage{orcidlink}
\usepackage{graphicx}
\usepackage{dcolumn}
\usepackage{bm}
\usepackage{hyperref}
\hypersetup{ 
    pdfnewwindow=true,      
    colorlinks=true,       
    allcolors=[RGB]{31 119 180}
} 
\usepackage{cleveref}
\usepackage{dsfont}
\usepackage{placeins}
\usepackage{todonotes}
\usepackage[utf8]{inputenc}
\usepackage{gensymb}
\usepackage{multirow}
\usepackage{amsmath}
\usepackage{amsfonts}
\usepackage{amssymb}
\usepackage{enumitem,amssymb}
\usepackage{array}   
\newcolumntype{L}{>{$}l<{$}} 
\newcolumntype{R}{>{$}r<{$}}
\newcolumntype{C}{>{$}c<{$}}
\usepackage{xspace}
\usepackage{braket}
\usepackage{units}
\usepackage{slashed}
\usepackage[normalem]{ulem}
\usepackage[format=plain,justification=RaggedRight,singlelinecheck=false]{caption}
\usepackage[format=plain,justification=centering,singlelinecheck=false]{subcaption}
\usepackage{tabularx}
\usepackage{nameref}

\usepackage{xpatch}
\makeatletter
\xpatchcmd{\@ssect@ltx}{\@xsect}{\protected@edef\@currentlabelname{#8}\@xsect}{}{}
\xpatchcmd{\@sect@ltx}{\@xsect}{\protected@edef\@currentlabelname{#8}\@xsect}{}{}
\makeatother
\usepackage{hyperref}

\graphicspath{{figures/}}

\let\Im\relax

\DeclareMathOperator{\Im}{Im}

\newcommand{\jpsi}{\ensuremath{J/\psi}\xspace}

\newcommand{\eg}{{\it e.g.}\xspace}

\newcommand{\ie}{{\it i.e.}\xspace}

\newcommand{\mevnospace}{\ensuremath{{\mathrm{\,Me\kern -0.1em V}}}}
\newcommand{\gevnospace}{\ensuremath{{\mathrm{\,Ge\kern -0.1em V}}}}
\newcommand{\tevnospace}{\ensuremath{{\mathrm{\,Te\kern -0.1em V}}}}
\newcommand{\mbnospace}{\ensuremath{{\mathrm{\,mb\kern -0.1em V}}}}

\newcommand{\mb}{\ensuremath{\mathrm{\,mb}}}
\newcommand{\nb}{\ensuremath{\mathrm{\,nb}}}
\newcommand{\fm}{\ensuremath{\mathrm{\,fm}}}
\newcommand{\mev}{\mevnospace\xspace}
\newcommand{\gev}{\gevnospace\xspace}

\newlist{todolist}{itemize}{2}
\setlist[todolist]{label=$\square$}
\usepackage{pifont}

\newcommand{\hallc}{$\jpsi$--007\xspace}

\newcommand{\sth}{\ensuremath{s_\text{th}}}
\newcommand{\E}[1]{\ensuremath{\times10^{#1}}}
\renewcommand{\th}{\ensuremath{\text{th}}}
\newcommand{\VMD}{\ensuremath{\text{VMD}}}
\usepackage{mathtools}

\newcommand{\AGH}{AGH University of Krakow, Faculty of Physics and Applied Computer Science, PL-30-059 Krak\'ow, Poland}
\newcommand{\catania}{INFN Sezione di Catania, I-95123 Catania, Italy}
\newcommand{\ceem}{Center for  Exploration  of  Energy  and  Matter, Indiana  University, Bloomington,  IN  47403,  USA}
\newcommand{\icn}{Instituto de Ciencias Nucleares,
Universidad Nacional Aut\'onoma de M\'exico, Ciudad de M\'exico 04510, Mexico}
\newcommand{\ific}{Instituto de F\'isica Corpuscular (IFIC), Centro Mixto CSIC-Universidad de Valencia, E-46071 Valencia, Spain}
\newcommand{\ifj}{Institute of Nuclear Physics, Polish Academy of Sciences, 31-342 Krak\'ow, Poland}
\newcommand{\indiana}{Department of Physics,
Indiana  University, Bloomington,  IN  47405,  USA}
\newcommand{\jlab}{Theory Center, Thomas  Jefferson  National  Accelerator  Facility, Newport  News,  VA  23606,  USA}
\newcommand{\ut}{Institute for Theoretical Physics, T\"ubingen University, D-72076 T\"ubingen, Germany}
\newcommand{\ur}{Institute for Theoretical Physics, Regensburg University, D-93040 Regensburg, Germany}
\newcommand{\messina}{Dipartimento di Scienze Matematiche e Informatiche, Scienze Fisiche e Scienze della Terra,
Universit\`a degli Studi di Messina, I-98166 Messina, Italy}
\newcommand{\scnuIQM}{Guangdong Provincial Key Laboratory of Nuclear Science, Institute of Quantum Matter, South China Normal University, Guangzhou 510006, China}
\newcommand{\scnuJLQM}{Guangdong-Hong Kong Joint Laboratory of Quantum Matter, Southern Nuclear Science Computing Center, South China Normal University, Guangzhou 510006, China}
\newcommand{\ub}{Departament de F\'isica Qu\`antica i Astrof\'isica and Institut de Ci\`encies del Cosmos, Universitat de Barcelona, E-08028, Spain}
\newcommand{\ucm}{Departamento de F\'isica Te\'orica, Universidad Complutense de Madrid and IPARCOS, E-28040 Madrid, Spain}
\newcommand{\uned}{Departamento de F\'isica Interdisciplinar, Universidad Nacional de Educaci\'on a Distancia (UNED), Madrid E-28040, Spain}
\newcommand{\ujk}{Institute of Physics, Jan Kochanowski University, PL-25-406  Kielce,  Poland}

\allowdisplaybreaks

\begin{document}

\preprint{JLAB-THY-23-3802}

\title{Dynamics in near-threshold \jpsi photoproduction}
\author{D.~\surname{Winney}\orcidlink{0000-0002-8076-243X}}
\email{dwinney@scnu.edu.cn}
\affiliation{\scnuIQM}
\affiliation{\scnuJLQM}
\author{C.~\surname{Fern\'andez-Ram\'irez}\orcidlink{0000-0001-8979-5660}}
\email{cesar@jlab.org}
\affiliation{\uned}
\affiliation{\icn}
\author{A.~\surname{Pilloni}\orcidlink{0000-0003-4257-0928}}
\email{alessandro.pilloni@unime.it}
\affiliation{\messina}
\affiliation{\catania}
 \author{A.~N.~\surname{Hiller Blin}\orcidlink{0000-0002-6854-6259}}
\affiliation{\ur}
\affiliation{\ut}
\author{M.~\surname{Albaladejo}\orcidlink{0000-0001-7340-9235}}
\affiliation{\ific}
\author{\L.~\surname{Bibrzycki}\orcidlink{0000-0002-6117-4894}}
\affiliation{\AGH}
\author{N.~\surname{Hammoud}\orcidlink{0000-0002-8395-0647}}
\affiliation{\ifj}
\author{J.~\surname{Liao}\orcidlink{0000-0003-1971-8787}}
\affiliation{\ceem}
\affiliation{\indiana}
\author{V.~\surname{Mathieu}\orcidlink{0000-0003-4955-3311}}
\affiliation{\ub}
\affiliation{\ucm}
\author{G.~\surname{Monta\~na}\orcidlink{0000-0001-8093-6682}}
\affiliation{\jlab}
\author{R.~J.~\surname{Perry}\orcidlink{0000-0002-2954-5050}}
\affiliation{\ub}
\author{V.~\surname{Shastry}\orcidlink{0000-0003-1296-8468}}
\affiliation{\ceem}
\affiliation{\indiana}
\affiliation{\ujk}
\author{W.~A.~\surname{Smith}\orcidlink{0009-0001-3244-6889}}
\affiliation{\ceem}
\affiliation{\indiana}
\author{A.~P.~\surname{Szczepaniak}\orcidlink{0000-0002-4156-5492}}
\affiliation{\ceem}
\affiliation{\indiana}
\affiliation{\jlab}

\collaboration{Joint Physics Analysis Center}

\begin{abstract}
The study of $\jpsi$ photoproduction at low energies has consequences for the understanding of multiple aspects of nonperturbative QCD, ranging from mechanical properties of the proton, to the binding inside nuclei, and the existence of hidden-charm pentaquarks. Factorization of the photon-$c\bar c$ and nucleon dynamics or Vector Meson Dominance are often invoked to justify these studies. Alternatively, open-charm intermediate states have been proposed as the dominant mechanism underlying \jpsi photoproduction. As the latter violates this factorization, it is important to estimate the relevance of such contributions.    
We analyse the latest differential and integrated photoproduction cross sections from the GlueX and \hallc experiments. 
We show that the data can be adequately described by a small number of partial waves, which we parameterize with generic models enforcing low-energy unitarity. 
The results suggest a nonnegligible contribution from open-charm intermediate states. Furthermore, most of the models present an elastic scattering length incompatible with previous extractions based on Vector Meson Dominance, and thus call into question its applicability to heavy mesons. Our results indicate a wide array of physics possibilities that are compatible with present data and need to be disentangled.
\end{abstract}

\date{\today}
\maketitle

\section{Introduction}

\begin{figure*}
    \centering
    \includegraphics[width=0.9\textwidth]{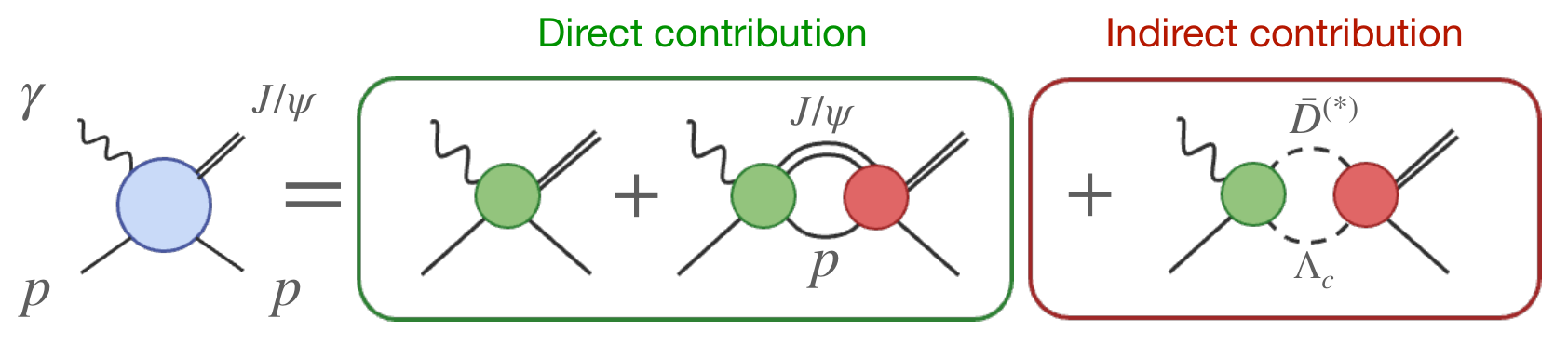}
    \caption{\label{fig:diagram} 
    Diagrammatic representation of the amplitudes in  \cref{eq:amp,eq:ampsplit}. Each PW amplitude $F_\ell^{\psi p}$ (blue) receives contributions from a (short-range) production coupling $f_\ell$ (green) as well as terms proportional to the loop function $G$ (Chew-Mandelstam phase space) and hadronic rescattering amplitude $T_\ell$ (red). The latter is summed over all intermediate channels that contribute.}
\end{figure*}
The photoproduction of charmonia near threshold has garnered substantial interest as it may give insight into a broad range of physics phenomena.
Since the charm quark mass is heavy, it has been argued 
 that charm production is a hard process. This motivates interpreting the amplitude in terms of factorized subprocesses, \ie a hard photon-$c\bar{c}$ conversion and a soft proton matrix element. The two subprocesses exchange dominantly gluons, as the exchange of charm is suppressed by the heavy quark mass, and the exchange of light quarks is OZI-suppressed~\cite{Brodsky:2000zc}. In this form, the photoproduction amplitude gives information about the internal structure of the proton and has been related to gluonic PDFs~\cite{Kharzeev:1998bz} or GPDs~\cite{Guo:2021ibg}, from which one extracts the gravitational form factors~\cite{Mamo:2019mka,Guo:2021ibg,Mamo:2022eui}, the trace anomaly contribution to the proton mass~\cite{Hatta:2018ina,Wang:2019mza,Ji:2021mtz,Han:2022btd}, and the mass radius~\cite{Wang:2021dis,Kharzeev:2021qkd,Mamo:2021krl}. 

 Extracting the elastic $\jpsi \, p$ amplitude is a necessary intermediate step for the determination of these quantities in some of frameworks, which is commonly done assuming Vector Meson Dominance (VMD)~\cite{Gryniuk:2016mpk,Strakovsky:2019bev,Pentchev:2020kao,Wang:2022xpw}. The elastic scattering process is also interesting \textit{per se}, as the small size of the $c\bar{c}$ pair compared to the nucleon suggests that the elastic scattering is driven by gluonic van der Waals forces and can be described using the  QCD multipole expansion~\cite{Luke:1992tm,Fujii:1999xn,Brodsky:1997gh}.
Close to threshold, where the relative momentum between proton and \jpsi is small, the interaction is expected to be attractive, and speculated to be strong enough to bind \jpsi to nucleons or even nuclei~\cite{Brodsky:1989jd,TarrusCastella:2018php}. The \jpsi-nucleon total cross section is also of interest for heavy ion collisions as final states with charmonia are a potential smoking gun for quark-gluon plasma~\cite{Barnes:2003vt,Rapp:2009my,Chao:1993ah}.

The discovery of hidden-charm pentaquarks in the $\jpsi \, p$ spectrum at LHCb~\cite{LHCb:2015yax,LHCb:2016ztz,LHCb:2019kea} has generated much interest in photoproduction searches, both theoretically~\cite{Wang:2015jsa,Kubarovsky:2015aaa,Karliner:2015voa,Blin:2016dlf,Winney:2019edt,Paryev:2022wov} and experimentally~\cite{GlueX:2019mkq,Meziani:2016lhg,SBS:2018,claspentaquark,Joosten:2018gyo}. Many theoretical studies highlight the role of open-charm channels in the formation of pentaquark signals~\cite{Wu:2010jy,Liu:2019tjn,Fernandez-Ramirez:2019koa,Du:2019pij}, which suggest these contributions may also be relevant in near-threshold photoproduction~\cite{Du:2020bqj}, and potentially break  factorization between hard charmonium production in the ``top" vertex and the soft nucleon recoil in the ``bottom" vertex.

Here we aim to address these questions by considering the photoproduction amplitude in a generic form, minimizing the model dependence and determining physical amplitude parameters solely from data.  We describe data using a small number of  $s$-channel partial waves (PWs), which we parameterize to satisfy unitarity constraints. This allows us to study the relevance of intermediate open-charm channels and test the VMD hypothesis. This approach is also general enough that resonance poles can emerge if data require them, allowing us to search for pentaquark states in the near-threshold region.

We consider the most recent data on total and differential cross sections from Jefferson Lab, in particular from GlueX~\cite{Adhikari:2023fcr} and the \hallc experiment in Hall~C~\cite{Duran:2022xag}. The interplay between the different production mechanisms is subtle, and these new data offer the possibility to discern the dynamics with more detail.

The rest of the paper is organized as follows. In \cref{S:formalism}, we review the unitary formalism to describe \jpsi production based on the near-threshold expansion. We consider four models of increasing complexity that offer different dynamical pictures and allow us to gauge systematic uncertainties. In \cref{S:results} we describe fits to the data and discuss implications for the nature of the underlying interactions. The four models describe the data with similar quality, and in some cases we find potentially large violations of factorization and VMD. Finally, in \cref{S:summary} we summarize our results and discuss future experimental measurements needed to confirm these  findings.

\section{Amplitude parameterization}
\label{S:formalism}
We consider the process $\gamma p \to \jpsi \,p$ in the region from threshold ($E_\gamma \simeq 8.2\gev$) to $12\gev$. The reaction amplitude depends on the standard Mandelstam variables $s$ and $t$, \ie the square of the center-of-mass energy and momentum transfer, respectively. In general, the amplitude also depends on the helicities of all four particles, but in the absence of polarization information the angular behavior can only be associated with the orbital motion determined by the angular momentum $\ell$, and there is little point in considering spin degrees of freedom at this stage. We thus approximate the four particles as spinless and write the unpolarized cross section in the usual form:
    \begin{equation}
        \frac{d\sigma}{dt} = \frac{1}{16\pi \, (s - m_p^2)^2} \, |F(s,t)|^2 ~,
    \end{equation}
and expand the scattering amplitude in terms of its $s$-channel PWs:
    \begin{equation}
        \label{eq:T-PWA}
        F\!\left(s, t\right) = \sum_\ell (2\ell+1) \, P_\ell\left(\cos\theta\right) \, F_\ell(s) ~,
    \end{equation}
where $\cos\theta \equiv \cos\theta(s,t)$ is the $s$-channel scattering angle. This expansion is particularly suitable to describe the region near threshold, where the infinite sum of partial waves is restricted by the angular momentum barrier factor, and is therefore expected to be saturated by a small number of terms. Furthermore, unitarity can be used to relate the imaginary part of the photoproduction amplitude to the hadronic final state interactions. In practice, unitarity is imposed effectively by considering only the most relevant two-body intermediate states. Thus we write: 
\begin{subequations}
 \label{eq:unitarity}
\begin{align}
 \Im F_\ell(s) &= F_\ell(s) \, \rho(s) \,T^\dagger_\ell(s)~,\\
 \Im T_\ell(s) &= T_\ell(s) \, \rho(s) \, T^\dagger_\ell(s)~,
 \end{align}
 \end{subequations}
 where $\rho$ is the two-body phase space of the intermediate state. When considering coupled channels, \cref{eq:unitarity} represent matrix equations, with the matrix elements $F^i_{\ell}(s)$ and $T_\ell^{ij}(s)$ corresponding to the photoproduction amplitude of the $i$-th final state and the $i\to j$ hadronic scattering amplitude, respectively. Although the $\bar{D}^{(*)}\Sigma_c^{(*)}$ channels have been proposed as relevant to the formation of hidden-charm pentaquarks~\cite{Wu:2010jy,Liu:2019tjn,Fernandez-Ramirez:2019koa,Du:2019pij}, recent GlueX data show no obvious structures at the corresponding thresholds. Instead, a dip at $E_\gamma \simeq 9\gev$ is observed with an estimated significance of 2.6$\sigma$,\footnote{The significance is only 1.4$\sigma$ when considering the probability of any two adjacent points having a similar significance.} and thus we rather consider the effect of the $\bar{D}^{(*)}\Lambda_c$ channels, whose thresholds are located at $E_\gamma \simeq 8.7$ and $9.4\gev$ respectively, as suggested in Ref~\cite{Du:2020bqj}. Since the data are available only  for the $\gamma p \to \jpsi \,p$ process, and not for open-charm final states, their effects enter only indirectly through rescattering. In order to limit the number of free parameters, coupled channels are implemented in the $S$-wave only, as threshold cusps are suppressed in higher waves, making it harder to disentangle the individual contributions of the various channels.  
 
A solution of \cref{eq:unitarity} is given by 
\begin{subequations}
 \label{eq:amp}
\begin{align}
 F_\ell(s) &= f_\ell\left(1 + G \, T_\ell\right) = f_\ell\left(1 - G \, K_\ell\right)^{-1}
 ~, 
 \\
 T_\ell(s) &= K_\ell \, (1 - G \, K_\ell)^{-1}~, \label{eq:ampb}
\end{align}
\end{subequations}
where the constraint of unitarity is satisfied as long as the $K$-matrix $K_\ell$ and the production vector $f_\ell$ are real in the physical region. The relation between amplitudes in \cref{eq:amp} is shown diagrammatically in \cref{fig:diagram}. We set the Chew-Mandelstam phase space $G = \delta^{ij} \, G_i$ to satisfy  $\Im G_i = \rho_i = q_i/8\pi\sqrt{s}$ and $G_i(s_i) = 0$ at the threshold of the $i$th intermediate state, $s_i = (m_{1i} + m_{2i})^2$~\cite{Wilson:2014cna}:
    \begin{align}
        \label{G}
        G_i &= \frac{s-s_i}{\pi}\int_{s_i}^\infty ds' \frac{\rho_i(s')}{(s'-s_i)(s'-s)}  \\
        &=-\frac{1}{\pi}\left[\rho_i \, \log\left(\frac{\xi_i + \rho_i}{\xi_i - \rho_i}\right) - \xi_i \, \frac{m_{2i} - m_{1i} }{m_{2i} + m_{1i}} \log\frac{m_{2i}}{m_{1i}} \right] \,.
        \nonumber
    \end{align}
Here $q_i = \lambda^{1/2}(s, m_{1i}^2, m_{2i}^2)/2\sqrt{s}$ is the intermediate state \mbox{3-momentum} and $\xi_i \equiv  (1 - s_i /s)/16\pi$.
For the (coupled channel) $S$-wave, we parameterize the production vector as a   constant, while keeping terms up to $O(q_i^2)$ in the low-energy expansion of the $K$-matrix, 
    \begin{equation}
        \label{eq:amp-K}
     f_S^i = n^i_S \quad \text{and}  \quad     K_S^{ij} = \alpha_S^{ij} + \beta_S^{i} \, q_i^2 \, \delta_{ij} ~,
    \end{equation}
with $\alpha_S^{ij}= \alpha_S^{ji}$  
due to  time reversal invariance. 
 We found that adding more terms 
  to the momentum expansion does not 
   improve the quality of the fits.  
For waves with $\ell\ge 1$, we consider only the  single  $\jpsi \,p$ channel and therefore drop the channel indices. In this case both the production vector and the $K$-matrix are parameterized as constants, $n_l$ and $\alpha_l$ respectively,  multiplied by the appropriate barrier factors: 
\begin{equation}
    \label{eq:f_and_K}
    f_\ell = (pq)^\ell \, n_\ell \quad \text{and} \quad K_\ell = q^{2\ell} \, \alpha_\ell ~. 
\end{equation}
Here $p = (s - m_p^2)/2\sqrt{s}$ is the incoming 3-momentum and $q\equiv q_{\psi p}$, as defined before.

In order to assess whether the current data can constrain the role of coupled channels, we consider three parameterizations of the $S$-wave amplitude:
    \begin{enumerate}
        \item \textbf{Single channel (1C):} Only interactions involving the $\jpsi \,p$ are included;
        \item \textbf{Two channels (2C):} 
        We include contributions from an intermediate $\bar D^* \Lambda_c$ channel;\footnote{The amplitude involving only the $\jpsi \,p$ and $\bar{D}\Lambda_c$ channels was also considered but found to not be significant.}
        \item \textbf{Three channels (3C):}  We include both $\bar{D}^{(*)}\Lambda_c$ channels. In this case we find two classes of solutions which we discuss separately below.  
           \end{enumerate}
Here the 1C parameterization is favored by the factorization picture of $\jpsi$ photoproduction, as charm exchanges are suppressed by the heavy quark mass and the amplitude can be decomposed into a ``top'' vertex involving the photon interaction with a $c\bar{c}$ pair and a ``bottom'' vertex that depends on proton structure.
Furthermore, it has been argued that near threshold the process is dominated by at most spin-2 exchanges in the $t$-channel~\cite{Ji:2021mtz}. This allows one to relate the $\jpsi \, p$ photoproduction amplitude to the gluonic component of the nucleon energy-momentum tensor. Fixed spin $t$-channel exchanges lead to an analytical dependence on $s$, and thus are not compatible with threshold cusps. 
The 1C case will be used at the base model with respect to which we evaluate the significance of extra thresholds.

Reference~\cite{Du:2020bqj} estimates that the production rates of open-charm systems are much larger than hidden charm, and thus the coupling to intermediate open-charm states dominates the process of interest. 
This is  representative of the 3C model in particular, if the production parameters for the $\jpsi\, p$ system are small relative to those describing the open-charm channels. We consider the 2C model as an intermediate case between these two parameterizations in which minimal freedom has been added to the 1C case to try to accommodate the apparent features of the total cross section. 

In order to have a comparable number of free parameters in three parameterizations of the $S$-wave, we consider both terms in the $K$-matrix (\ie $\alpha_S^{ij}$ and $\beta_S^{i}$) in the 1C and 2C cases, while the 3C parameterization keeps only the constant term in all channels. We find an adequate description of the angular dependence when truncating to $\ell_\text{max} = 3$ in all cases.   

Even if no explicit $K$-matrix pole is included, the amplitude in \cref{eq:amp} can produce poles in the complex energy plane in all three parameterizations. 
If the pole appears sufficiently close to the physical region, it can be interpreted as a signal of a hidden-charm pentaquark.  

    \begin{figure*}
        \centering
        \includegraphics[width=\textwidth]{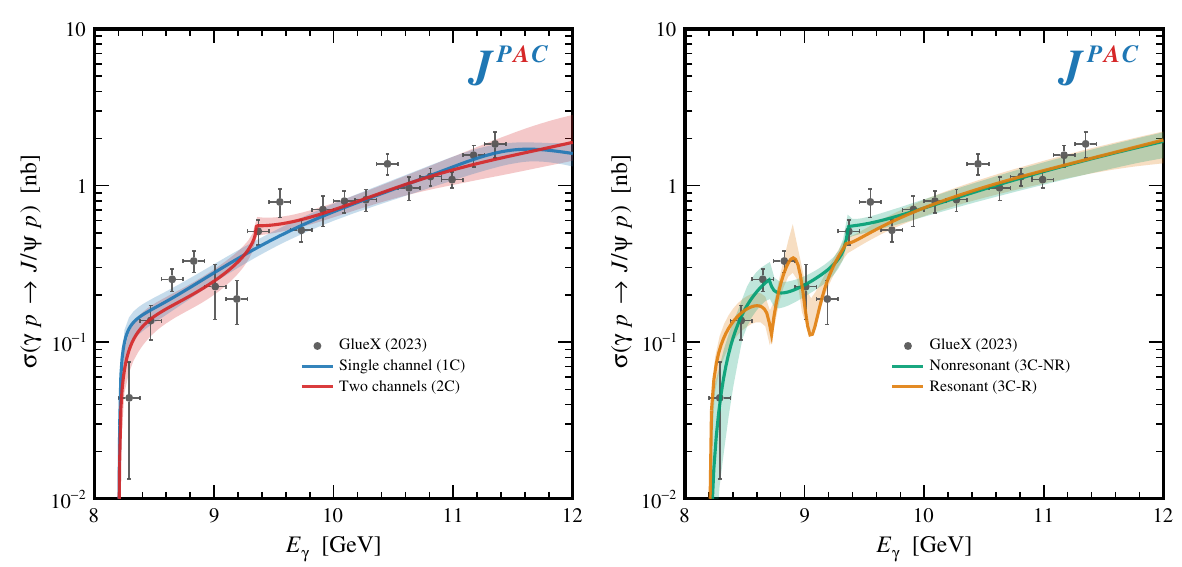}
        \caption{Fit results for the integrated cross section of all four models compared to GlueX data from~\cite{Adhikari:2023fcr}. Bands correspond to 1$\sigma$ uncertainties from bootstrap analysis.}
        \label{fig:integrated_bs}
    \end{figure*}
\section{Results and discussion}
\label{S:results}
We analyze all of the recent Jefferson Lab data from the GlueX~\cite{Adhikari:2023fcr} and \hallc~\cite{Duran:2022xag} experiments. This covers both the integrated cross section for the photon energies $E_\gamma = 8.2$--$11.4\gev$ and differential cross section reported in 15 energy bins.  
The differential cross section measurements of  GlueX cover the entire physical $t$ range. In total we have $142$ data points that we fit with each model  described in \cref{S:formalism}. We fit differential data at the reported $\left\langle t\right\rangle$ and $\left\langle E_\gamma \right\rangle$ values, which are averaged over the bin. Fits are performed  by minimizing the standard $\chi^2$ function with  the experimental statistical and (uncorrelated) systematic uncertainties added in quadrature. We ignore correlations between integrated and differential cross sections. We note that the datasets from the two experiments have different normalization uncertainties ($\simeq 20\%$ for GlueX and $\simeq 4\%$ for \hallc). The fits are conducted under the assumption that the two datasets are consistent, and the correlated normalization errors are later included in the error analysis. The fit parameters were all initialized randomly and fits were repeated to sufficiently probe the parameter space.
The resulting parameters for the best fits are summarized in \cref{app:errors}. For the 3C parametrization we find two qualitatively different fit results with similar $\chi^2$: one with a pole near the real axis which we label 3C-R (resonant), and one without a nearby pole, which we label 3C-NR (nonresonant).

\begin{figure*}
        \centering
        \includegraphics[width=\columnwidth]{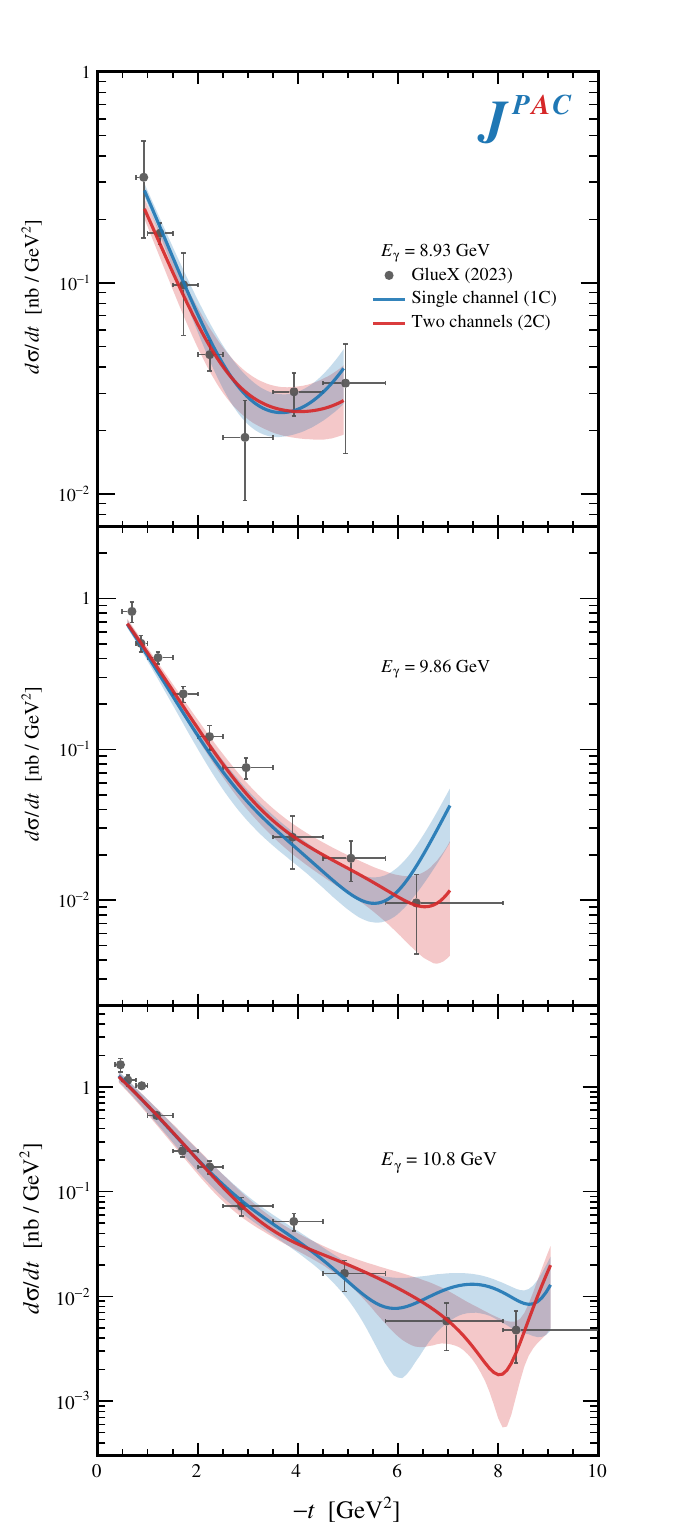}
        \includegraphics[width=\columnwidth]{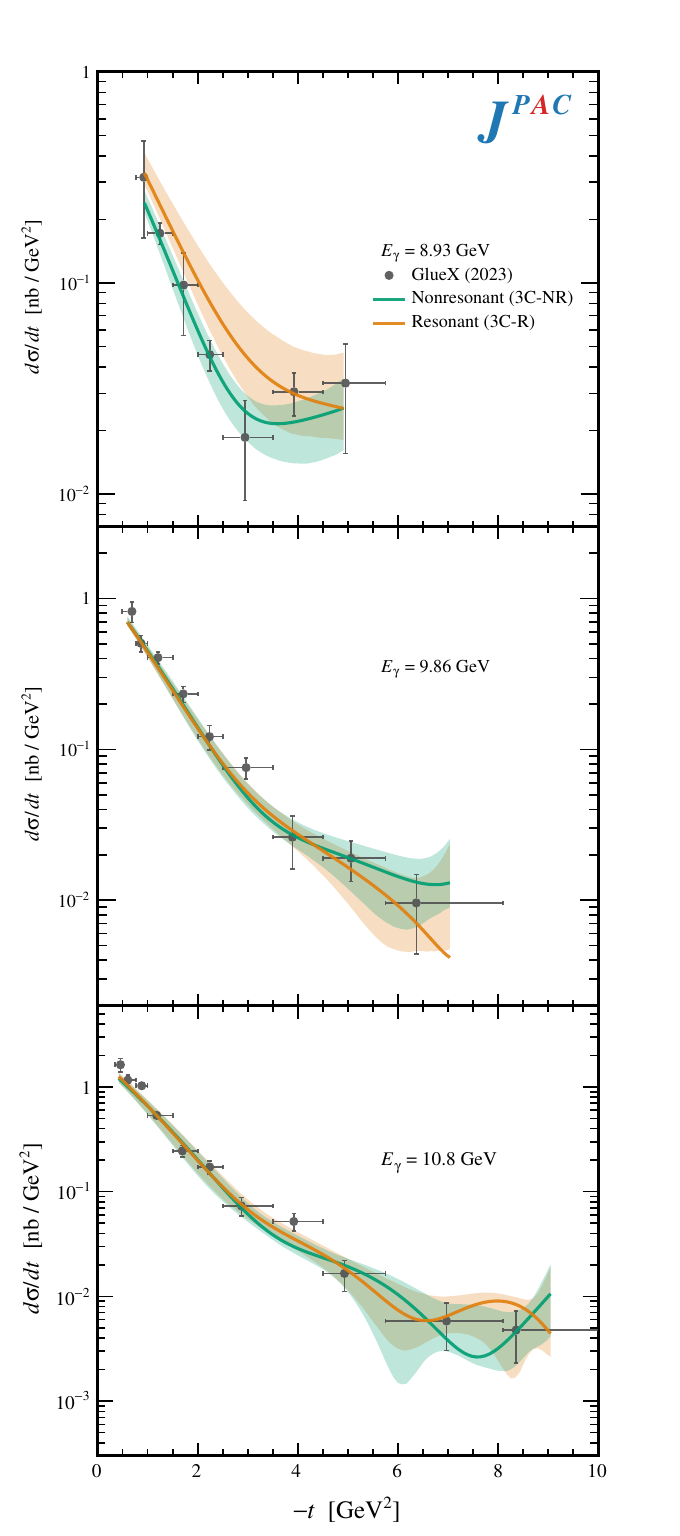}
        \caption{Fit results for the differential cross section of all four models compared to GlueX data from~\cite{Adhikari:2023fcr}. The bands correspond to the $1\sigma$ uncertainties from the bootstrap analysis. The range of the data represents the entire physical $t$ range at each fixed energy. The plotted theory curves are calculated at the average $\left\langle E_\gamma\right\rangle$ for each experimental energy bin.}
        \label{fig:differential_bs}
    \end{figure*}

\begin{figure*}
        \centering
        \includegraphics[width=\columnwidth]{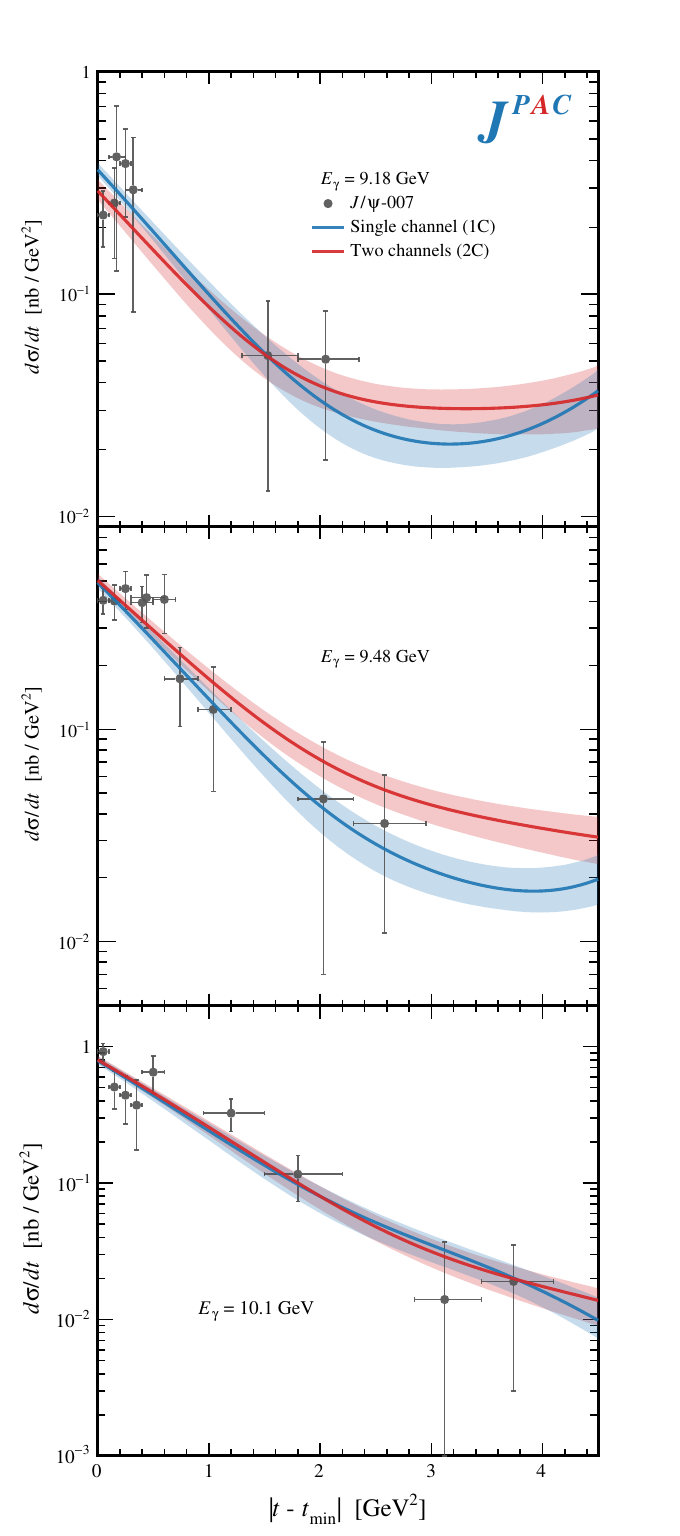}
        \includegraphics[width=\columnwidth]{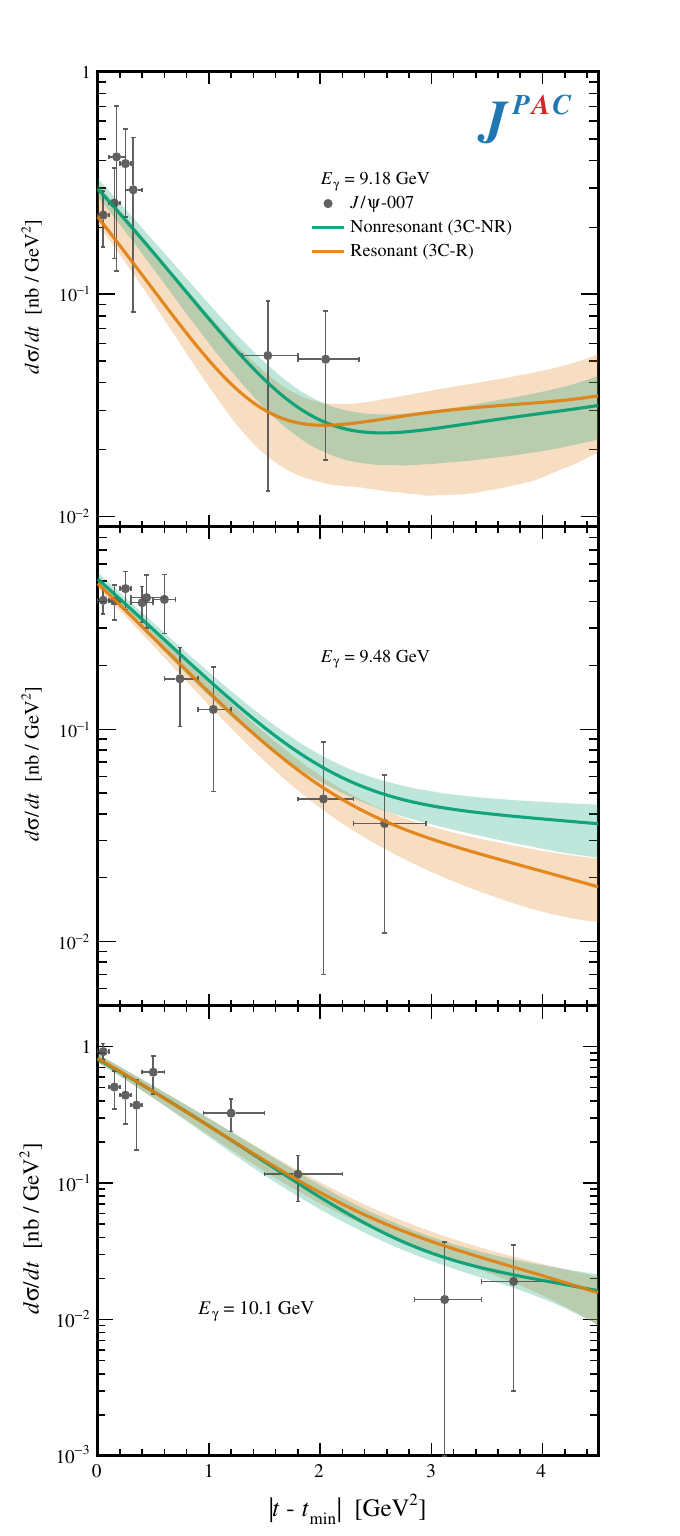}
        \caption{Fit results for the differential cross section for all four models compared to \hallc data from~\cite{Duran:2022xag}. Here we only show three characteristic example slices in between those shown in \cref{fig:differential_bs}, but all data are included in the fits. The bands correspond to the $1\sigma$ uncertainties from the bootstrap analysis. The plotted theory curves are calculated at the average $\left\langle E_\gamma\right\rangle$ for each experimental energy bin.}
        \label{fig:differential_bs2}
    \end{figure*}

We determined uncertainties using a bootstrap approach~\cite{JPAC:2021rxu}, taking into account the  statistical, systematic, and normalization uncertainties by assuming they are normally distributed. Further discussion of the propagation of uncertainties can be found in \cref{app:errors}. 
The resulting cross sections and associated uncertainties are shown in \cref{fig:integrated_bs,fig:differential_bs,fig:differential_bs2}. 


The simplest parameterization considered is the 1C scenario which exhibits a smooth energy behavior, since it contains no other channels that can give rise to threshold cusps. 
In the integrated cross section, the data points at the dip lie at least $2\sigma$ away from the fit curve, consistent with significance estimations in Ref.~\cite{Adhikari:2023fcr}.

Extensions of the $K$-matrix in \cref{eq:amp-K,eq:f_and_K} were considered to study the systematics of the 1C results. Additional $O(q^4)$ and $O(q^2)$ terms were added to the $S$- and $P$-waves, respectively, but yielded no significant improvement over the original fit. Higher waves beyond $\ell = 3$ were also considered, but had little impact on the best fit values of the lower PWs. 

Adding the $\bar{D}^*\Lambda_c$ channel in the 2C model leads to a clear threshold cusp around $E_\gamma \simeq 9.5\gev$. This improves the fit quality with respect to the single channel case, but the significance is not high enough to definitively favor this result over the 1C curve.
Repeating the same analysis considering the lighter $\bar{D}\Lambda_c$ channel instead does not constitute a significant improvement over the 1C fit, with best fit open-charm parameters found to be consistent with zero. 

The curves with the most structure arise from the 3C model containing both open-charm thresholds. These fits showcase dips similar to those apparent in the data but differ in the precise lineshape between the two open-charm thresholds, meaning comparison between these is sensitive to the details of the cusp-dip structure. We see the 3C-NR shape is reminiscent of predictions from the box diagram calculation in~\cite{Du:2020bqj}. The other fit, labeled \mbox{3C-R}, exhibits an even  more pronounced lineshape which is due to the presence of a nearby pole singularity, the implications of which will be discussed in \cref{sec:pc}. 

\subsection{Momentum transfer distributions}
\label{sec:tdist}

Differential data near threshold have been proposed as a means to access the gluonic gravitational form factors and mass radius of the proton (see \eg~\cite{Ji:2021mtz,Kharzeev:2021qkd,Duran:2022xag} and references therein). The observed behavior in $t$ poses interesting questions about the $\jpsi \, p$ interaction itself. 

At high energies, the photoproduction of vector quarkonia has been extensively studied at HERA~\cite{ZEUS:2002wfj,H1:2005dtp}. In this energy region, the process is diffractive and generally understood through gluonic exchanges, realized \eg as a Pomeron~\cite{Donnachie:1998gm} or in a color-dipole model~\cite{Caldwell:2001ky}. The differential distributions are characterized by their ``diffractive peak'' at forward $t$ and exponential dropoff at high transferred momentum. 
Previous measurements of the differential cross section seem to observe the same behavior even at lower energies~\cite{Camerini:1975cy,GlueX:2019mkq},  
and are confirmed by the newest GlueX measurement for $E_\gamma > 9\gev$. At the lowest energy value $\left\langle E_\gamma\right\rangle = 8.9\gev$, the last few bins at largest $t$ seem to turn upward and have drawn attention as potential indications of $u$-channel exchanges or other $s$-channel contributions.  
However, since each PW series in the $s$-, $t$- or $u$-channel is a full representation of the amplitude, these contributions cannot be simply added, but the whole process has to be studied consistently from one perspective. 

\begin{figure*}
    \centering
    \includegraphics[width=\textwidth]{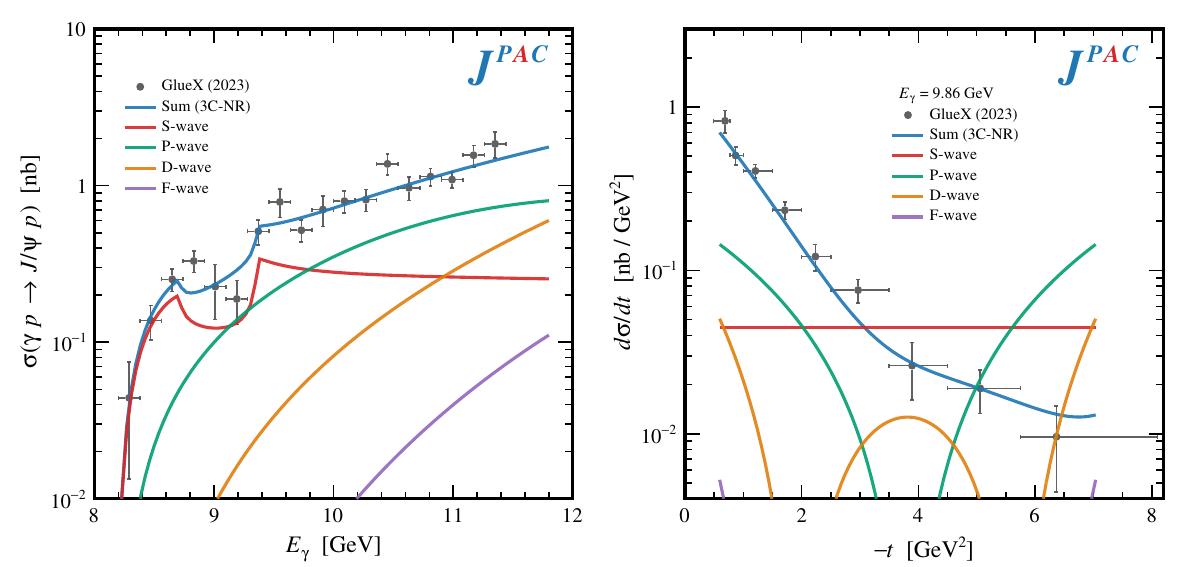}
    \caption{Comparison of individual PW contributions to the cross sections from GlueX~\cite{Adhikari:2023fcr} using the 3C-NR best fit parameters. In the differential cross section, the Legendre polynomials interfere to give rise to the sharp $t$ distribution reminiscent of the diffractive peak. A similar hierarchy of waves is seen in all fit results.}
    \label{fig:PW_compare}
\end{figure*}

Explaining the apparent exponential behavior of the $t$ distributions with the finite PW sum of \cref{eq:T-PWA} does not immediately seem natural.
Each $s$-channel PW has polynomial angular (and therefore $t$) dependence, unlike commonly used dipole or exponential form factors, so one would naively expect that a large number of PWs are needed to describe data. Instead, we find a good description only considering terms with $\ell \le 3$. The emergence of the sharp asymmetric $t$-distribution is due to interference between the PW amplitudes, as waves with odd and even $\ell$ interfere constructively at forward angles and destructively at backward ones (\ie  through $P_\ell(\cos\theta = \pm1) = (\pm1)^\ell$). Individual contributions to the cross sections are plotted in \cref{fig:PW_compare} for the 3C-NR case. 

In order to more quantitatively explore the convergence of the PW series, we may examine the  radius of interaction  $r$ which enters with the angular momentum barrier, 
    \begin{equation}
        \label{radius}
        r^{2\ell} \equiv \lim_{s\to\sth}  \left|\frac{F_\ell(s) / (pq)^\ell}{F_{S}(s)}\right|~,
    \end{equation}
where $s_\text{th} = (m_\psi + m_p)^2$ is the photoproduction threshold.
As long as $pq \, r^2 < 1$, we may expect any subsequent waves to be suppressed and the use of a finite number of PWs to be justified. Technically speaking, the interaction radius varies per PW, but we care about the typical value with which to characterize the rate of convergence. 
For all fit results, the radius is found to be $r \simeq 0.1 \fm$. Since the PWs may also vary independently as a function of $s$, we additionally consider the limit \cref{radius} taken to the end point energy of the data, $E_\gamma \simeq 12 \gev$. We find the energy dependence is extremely mild and we maintain the same average $r$ value. Thus extrapolating this to the transition energy satisfying $pq \,r^2 = 1$, we may expect the description in terms of $s$-channel PWs to hold up to about $E_\gamma \sim 14\gev$. At energies beyond this point, there is no suppression of higher waves and the infinite series must be resummed, characteristic of the Regge regime.

The fact that all amplitude  models reproduce the differential data accurately seems to suggest that the shape of the momentum transfer distribution alone does not discriminate details of the individual PWs with the current precision. Furthermore, \cref{fig:differential_bs} demonstrates that all models, resonant and nonresonant, reproduce the apparent upward behavior of the lowest energy slice of the GlueX measurement. This suggests the enhancement at backward $t$ very close to threshold is not necessarily indicative of $s$-channel resonances.

\subsection{Production mechanisms}
\label{sec:mech}
As previously mentioned, establishing 
 that charmonium photoproduction near threshold is due to short-range fluctuations in the photon beam is needed in order to be able to use this reaction to extract the proton tensor charge.  
If the contribution from charm exchange is found to be sizable, this process may be a low-energy probe of the intrinsic charm component of the nucleon wave function~\cite{Brodsky:1980pb,Saleev:1994dn}, whose extraction from inclusive measurements at higher energies has recently been studied~\cite{Ball:2022qks,Guzzi:2022rca}.

The formalism in \cref{S:formalism} allows us to clearly identify production quantities for each channel individually, and thus we can test the factorization hypotheses by assessing the strength of open charm contributions based on available data. 
The coupled-channel $S$-wave amplitude in \cref{eq:amp} can be explicitly written as 
\begin{widetext}
    \begin{align} \label{eq:ampsplit}
        F^{\psi p}_S(s) = 
        \overbrace{n_S^{\psi p}\left(1 + G^{\psi p} \, T_S^{\psi p,\psi p}\right)}^{\displaystyle F^{\psi p}_\text{direct}(s)}
        +\overbrace{\left( \,n_S^{\bar D \Lambda_c} \, G^{\bar D \Lambda_c} \, T_S^{\bar D \Lambda_c,\psi p}+ n_S^{\bar D^* \Lambda_c} \, G^{\bar D^* \Lambda_c} \,T_S^{\bar D^* \Lambda_c,\psi p}\right)}^{\displaystyle F^{\psi p}_\text{indirect}(s)} 
        ~,
    \end{align}
\end{widetext}
where we identify terms corresponding to the two production mechanisms as depicted diagrammatically in \cref{fig:diagram}. The ``direct" amplitude is composed of contributions in which the initial $\gamma p$ state couples directly to $\jpsi \, p$, while in the ``indirect'' contributions $\gamma p$ produces an intermediate open charm pair first, before rescattering into the final $\jpsi\,p$. 

The relative strengths of these two terms allow us to gauge which photoproduction mode is more relevant in the region of interest. To more easily quantify this, we define the ratio:
    \begin{equation}
        \label{eq:zeta}
        \zeta_\text{th} = \frac{ \left|F^{\psi p}_\text{direct}\!\left(s_\text{th}\right)\right| }{ \left|F^{\psi p}_\text{direct}\!\left(s_\text{th}\right)\right| + \left|F^{\psi p}_\text{indirect}\!\left(s_\text{th}\right)\right|}
        ~.
    \end{equation}
 Clearly the 1C case has $\zeta_\text{th} = 1$, while $\zeta_\text{th} \simeq 0$ would indicate the $\jpsi$ is almost entirely produced through intermediate open charm. Since the latter requires flavor exchange between the top and bottom vertices, it would explicitly break factorization. 

Examining the lineshapes in \cref{fig:integrated_bs}, one might naively assume that the cusps are a small contribution to an otherwise smooth background from the direct reaction. However, since the various terms are added at the amplitude level, this might not necessarily be the case. We tabulate the extracted values of $\zeta_\th$ in \cref{tab:results}, which indicates the indirect contributions play a non-negligible role. The direct amplitude contributes only $\zeta_\th \lesssim 75\%$ at 90\% confidence level in all coupled channel parameterizations, with the 3C-R result even compatible with $\zeta_\th = 0$. Clearly, deviations of $\zeta_\th$ from unity are due to the presence of the $9\gev$ dip in the data, which can only be captured with a sizable coupling to open charm. Thus if future data confirm the dip, the implications for factorization and the extraction of proton observables will need to be addressed.
    \begin{table*} 
        \centering
        \caption{\label{tab:results} Summary of fit results. For each solution we tabulate:
        the number of parameters, $\chi^2$, and reduced $\chi^2$. We also tabulate the 90\% CL interval of each 
        dynamical quantity described in the text (see \cref{app:errors}). In order these are: the ratio of production mechanisms defined in \cref{eq:zeta}, the VMD ratio in \cref{eq:RVMD} (extracted for both $\theta=0$ and $t=0$ prescriptions), and the elastic $\jpsi \,p$ scattering length in \cref{eq:scatteringlength}.
        }
        \begin{ruledtabular}
            \begin{tabular}{l c c c c}
                & \textbf{1C} & \textbf{2C} & \textbf{3C-NR} & \textbf{3C-R} \\ \hline
                Parameters     & $9$      & $13$      & $15$       & $15$      \\
                $\chi^2$     & $166$  & $144$  & $141$   & $143$  \\
                $\chi^2$/dof & $1.25$    & $1.12$    & $1.11$     & $1.13$ \\   \hline
                $\zeta_\text{th}$    & 1      & $[0.56,\, 0.74 ]$  & $[0.36,\, 0.63]$     & $[0.03,\, 0.62]$  \\ 
                $R_\text{VMD}(\theta = 0)$   & $[0.45,\, 0.73]\E{-2}$   & $[0.39,\, 1.62]\E{-2}$   & $[0.03,\, 1.74]\E{-2}$  & $[1.4\E{-2},\, 0.58]$  \\
                $R_\text{VMD}(t=0)$     & $[1.3,\,2.0]\E{-2}$ & $[1.3,\, 5.1]\E{-2}$ & $[0.08,\, 8.9]\E{-2}$ & $[5.4\E{-2}, 1.8] $ \\
                $a_{\psi p}$ [fm]      & $[0.56,\, 1.00]$ & $[0.11,\, 0.79]$   & $[ -2.77,\, 0.35]$ & $[-0.04,\, 0.19]$ \\ 
            \end{tabular}
        \end{ruledtabular}
    \end{table*}

\subsection{Vector meson dominance}
\label{S:VMD}
The VMD assumption has been used extensively in the analysis of photon-hadron interactions, and in particular in the extraction of proton observables~\cite{Kharzeev:1998bz,Kou:2021bez,Wang:2022tzw}. This posits that the photoproduction interaction can be modeled by replacing the incident photon by the hadron spectral function, generally modeled as a sum of vector meson propagators, and multiplying each term by a known constant related to the vector meson electromagnetic width.
Although it has been argued that, in particular for heavy quarkonia, the sum over higher vectors should be retained (so-called ``Generalized VMD''~\cite{Gari:1984rr}), most of the literature about charmonium photoproduction at threshold restricts the sum to the lightest $\jpsi$, as it allows one to relate photoproduction to the elastic scattering amplitude:
    \begin{align}
        \label{eq:vmd}
        F^{\psi p} (s,x) &=  g_{\gamma\psi} \, T^{\psi p, \psi p}(s,x)~.
    \end{align}
Here $x = t$ or $\theta$ depending on whether this relation is considered at fixed momentum transferred or scattering angle, \ie\ through two different $t = t(s,\cos\theta)$ for the photoproduction and elastic reactions.\footnote{For a more detailed discussion regarding these two forms of \cref{eq:vmd}, see Ref.~\cite{Pentchev:2020kao} and references therein.}  
The proportionality constant $g_{\gamma\psi}$ represents the \mbox{$\gamma \to c\bar{c}$} transition strength, and is related to the $\jpsi$ decay constant, $g_{\gamma\psi} = e\,f_\psi/m_\psi \simeq 0.0273$ extracted from the \jpsi electronic width. In quark models, the latter is related to the quarkonium wave function at the origin~\cite{VanRoyen:1967nq,Eichten:1995ch}. The core assumption is that the proton acts as a spectator when the \jpsi is formed and thus the energy dependence of production and elastic amplitudes is the same. 

In contrast, the structure of the photoproduction amplitude dictated by near-threshold unitarity in \cref{eq:amp} illustrates that photoproduction and elastic scattering amplitudes are not necessarily proportional. 
While it may be the case that VMD still holds, our analysis does not rely on it, and the relation between production and elastic amplitudes is determined solely by data. This means we may directly compare the photoproduction and elastic amplitudes and gauge if VMD is justified in the near-threshold region.
We quantify this test by defining the ratio:
    \begin{equation}
        \label{eq:RVMD}
        R_\VMD(x) = 
        \left| \frac{F^{\psi p}(\sth, x) \big / g_{\gamma\psi}}{T^{\psi p,\psi p}(\sth, x)}\right| ~,
    \end{equation}
where the numerator would be the elastic amplitude calculated assuming \cref{eq:vmd}, while the denominator is the one extracted directly from \cref{eq:amp}. We fix $s=\sth$ for concreteness and use the reference value of $g_{\gamma\psi}$ quoted above.  Since \cref{eq:T-PWA} is entirely analytic, we may compute $R_\text{VMD}$ either at fixed $\theta = 0$,\footnote{The same relation holds for any fixed value of $\theta$, we select $\theta=0$ for aesthetic reasons.} or at the unphysical point $t=0$ --- \eg as done in~\cite{Wang:2022xpw,Strakovsky:2019bev} or  in~\cite{Kharzeev:1998bz,Gryniuk:2016mpk,Kou:2021bez} respectively.  
If VMD is an accurate approximation of the production amplitude, we should expect $R_\text{VMD} \simeq O(1)$. Instead, the results in \cref{tab:results} suggest that VMD underestimates the amplitude by two orders of magnitude regardless of evaluation in almost all the fit results. The only exception is the 3C-R model which has the largest uncertainties. 

If these results were to be confirmed, the applicability of VMD in the heavy quarkonium sector would be severely questioned, affecting the widespread application of VMD in theoretical studies. For example, the current upper limits on hidden-charm pentaquark branching fractions in photoproduction are based on VMD models and sit at the sub-1\% level~\cite{GlueX:2019mkq,Winney:2019edt}. If VMD is so drastically violated, pentaquarks may still have sizable branching ratios, $\mathcal{B}(P_c \to \jpsi\,p) \simeq O(10\%)$, but a much smaller photocoupling than expected (compatible with estimations in~\cite{Ortiz-Pacheco:2018ccl}), which makes them more difficult to observe in photoproduction. 

\subsection{\boldmath $\jpsi \,p$ scattering length}
\label{sec:sl}

One immediate consequence of a failure of VMD is its effect on the extraction of the elastic scattering length from photoproduction data. This is of fundamental importance, as it may enter the proton mass decomposition~\cite{Kou:2021bez}, provides motivation for color transparency~\cite{Paryev:2015xoa} and suggests the possible emergence of bound states~\cite{Brodsky:1989jd,Kaidalov:1992hd,Eides:2017xnt}.

In our normalization, the scattering length $a_{\psi p}$, is related to the $S$-wave elastic scattering amplitude close to threshold by
    \begin{equation}
        \label{eq:scatteringlength}
        T_{S}^{\psi p , \psi p} 
         = \frac{8\pi \, \sqrt{s_\th}}{-a^{-1}_{\psi p} - i \, q} + O(q^2)
        ~.
    \end{equation}
Using VMD, \ie assuming \cref{eq:vmd}, this relation leads to:
\begin{align}
\text{VMD:~}& F^{\psi p} (\sth,x) =  -8\pi \sqrt{\sth}\, g_{\gamma\psi} \, a_{\psi p}~.
\end{align}
This means that the square of the scattering length is assumed to be directly related to the normalization of the photoproduction differential cross section at threshold (extrapolated at $t=0$), or the normalization of the total cross section divided by the phase space (for $\theta=0$). In this way, a small photoproduction cross section will directly translate into a small scattering length, which does not need to be the case.
In our framework, unitarity gives a relation at the PW level, which means that the photoproduction is related to the scattering one at fixed $\theta$. In the simplest 1C case, the equivalent expression using \cref{eq:unitarity,eq:scatteringlength} yields
\begin{align}
\text{1C:~}&
F^{\psi p} (s\to \sth,\theta) =  n_S^{\psi p} \left(1 - i \, q  \, a_{\psi p} \right)+ O(q^2)~, 
\end{align}
where one sees the scattering length drops out of the normalization and can only be extracted from the energy dependence. This relation gets even more complicated for coupled channels, as the indirect contributions of \cref{eq:ampsplit} will enter the equation, but the conclusion is the same: the normalization of photoproduction and the elastic scattering length are in general independent. 

Scattering lengths $O(1\fm)$ would indicate a typical hadronic interaction between the charmonium and nucleon, and are consistent with the range of theoretical predictions based on the QCD multipole expansion~\cite{Sibirtsev:2005ex,Krein:2020yor}, gluonic van der Waals forces~\cite{Brodsky:1997gh}, QCD sum rules~\cite{Hayashigaki:1998ey}, and some lattice QCD extractions~\cite{Sugiura:2019fue,Yokokawa:2006td}. Previous VMD-based extractions from data have yielded scattering lengths up to three orders of magnitude smaller~\cite{Pentchev:2020kao}, similar to expectations from effective field theories~\cite{Du:2020bqj} and some lattice studies~\cite{Skerbis:2018lew}. The smallness of the scattering length relative to the proton size has been argued to be related to the compact size of the $c\bar{c}$ pair rendering it ``transparent" to the proton.

The extracted values for all fits are reported in \cref{tab:results}. Fits 1C and 2C, which have the best constrained parameters, give scattering lengths of the order of a Fermi at 90\% CL, in stark contrast to VMD-based extractions. The values obtained from the 3C models, on the other hand, are consistent with zero in both the resonant and nonresonant models. Interestingly the 3C-NR interval reveals a propensity for larger, negative scattering lengths while the 3C-R extracts $|a_{\psi p}| \lesssim 0.2$ fm at a 90\% CL. 

While there is a clear preference for larger values of the scattering length and severe violation of VMD, the poorly constrained 3C models do not allow definitive conclusions to be drawn. Further data on the dip region and direct measurements of open charm photoproduction will better constrain the parameters of coupled channel models, and therefore resolve the size of the scattering length.

\subsection{\boldmath Total $\jpsi \, p$ cross section}

 Establishing a relation between charmonium photoproduction and elastic scattering is also of relevance for quantitative descriptions of the charmonium interaction and evolution within the many-body hadronic medium at the final stage of heavy ion collisions~\cite{Andronic:2015wma,Rapp:2008tf,Barnes:2003vt,Voloshin:2007dx,Hilbert:2007hc}. Phenomenological simulations of the charmonia suppression in these collisions, which have nontrivial implications as a signature of the quark-gluon plasma phase, would rely on accurate knowledge of such cross sections.  

Since the charmonium scattering is not achievable experimentally, estimations for cross sections must be inferred indirectly. Until recently, the lack of data on charmonium production near threshold meant that the cross section at the low energies was poorly known. The existing estimates from near-threshold photoproduction data came from SLAC in the 1970s and used either VMD assumptions~\cite{Camerini:1975cy} or the $A$-dependence considering various nuclear targets~\cite{Anderson:1976hi} to estimate the total cross section at a beam energy $E_\gamma \simeq 20 \gev$ ($\sqrt{s}\simeq 6.2 \gev$). The values extracted using the different methods, $\sim0.3 \mb$ and $\sim 4 \mb$ respectively, revealed a large discrepancy between the extractions with VMD yielding a significantly smaller value (see also discussion in Ref.~\cite{Hufner:1997jg}). 

 Theoretical estimates for the total cross section, \eg using color dipole models~\cite{Kharzeev:1994pz}, constituent quark models~\cite{Black:2002rs,Martins:1995nb}, or meson exchange models~\cite{Oh:2000qr,Liu:2001yx,Molina:2012mv} have also predicted a broad range of values for the cross section from fractions of a $\mb$ to upwards of $\sim10 \mb$  near threshold. 

\begin{figure}
    \centering
    \includegraphics[width=\columnwidth]{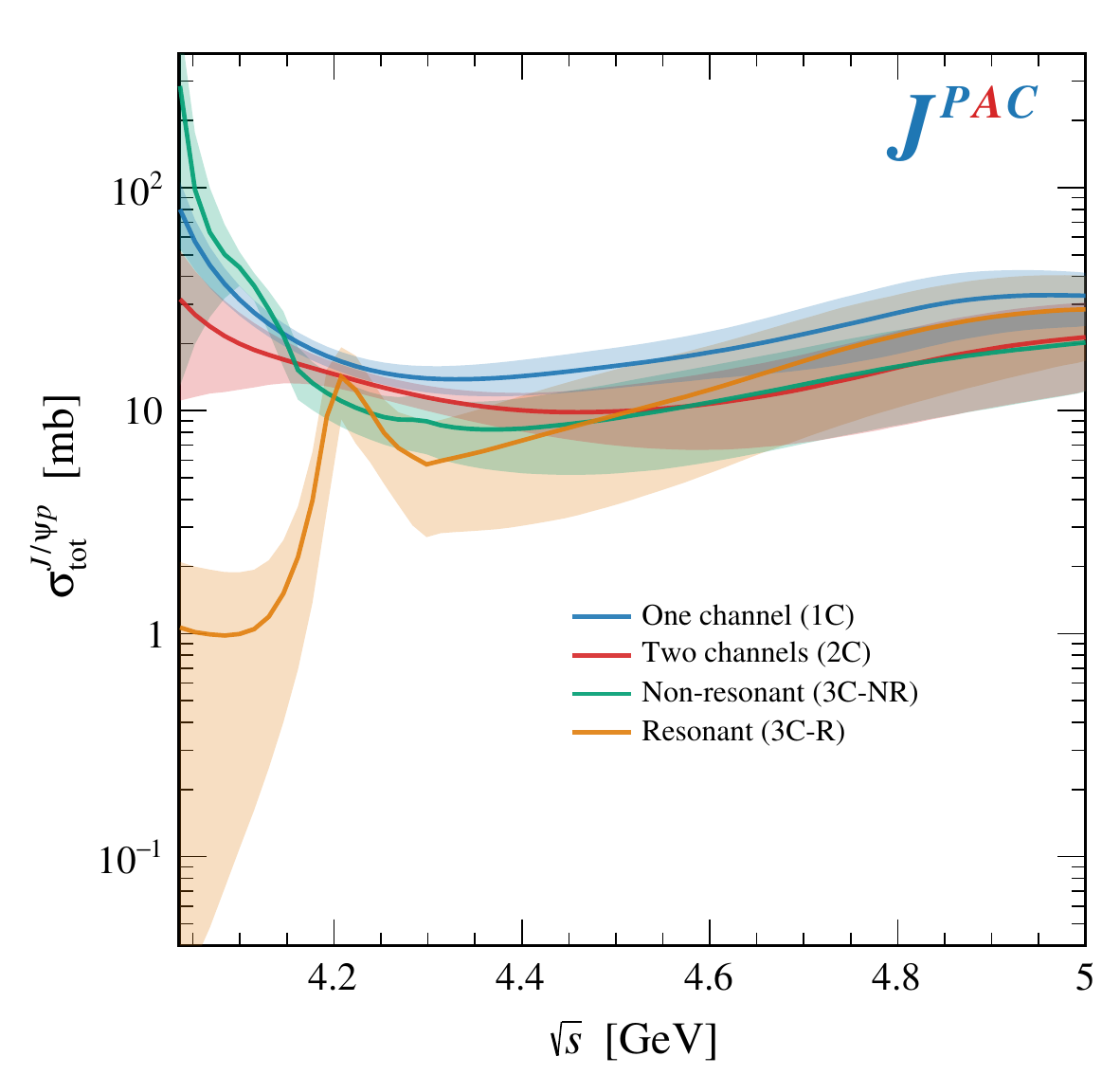}
    \caption{Near-threshold total hadron cross section of the charmonium-nucleon system calculated for each of our fit results. Bands correspond to $1\sigma$ uncertainties calculated by bootstrap analysis. The pentaquark contribution is evident in the resonant case.}
    \label{fig:sigmatot}
\end{figure}

 Because our formalism has access to the elastic amplitude directly, we may consider the total $J/\psi \,p $ cross section from our fit results. Using the optical theorem we calculate:
    \begin{equation}
        \label{eq:sigtot}
        \sigma_\text{tot}^{\psi p} = \frac{1}{2q\sqrt{s}} \, \Im T^{\psi p, \psi p}(s,t = 0)~,
    \end{equation}
which we plot for all fit cases in \cref{fig:sigmatot}. At energies just above threshold, the overall size of the cross section is dominated by the $S$-wave scattering length. As demonstrated in \cref{sec:sl}, this is sensitive to the dynamics of the $S$-wave and varies drastically depending on the parameterization used. Further we notice the clear resonant peak that appears in the 3C-R model. At higher energies, the cross section is dominated by higher waves where we see a closer overlap of values. At $\sqrt{s} = 5\gev$ we find $\sigma_\text{tot}^{\psi p} \gtrsim 8 \mb$  at a 90$\%$ CL in all amplitudes which include open-charm contributions, while the 1C case has $\gtrsim20\mb$.

These numbers are roughly compatible with the SLAC measurement not assuming VMD, although an explicit quantitative comparison is not possible with our near-threshold formalism, as the data are at energies beyond the radius of convergence of the PW expansion (see \cref{sec:tdist}).

\subsection{Pentaquark searches}
\label{sec:pc}

The parameterization of the $K$-matrix is general enough that poles may still emerge if favored by the data. In particular this allows for the possibility of pentaquark poles which feature non--Breit-Wigner lineshapes due to complicated multichannel dynamics in the $S$-wave. 

 For a given set of parameters, we may locate poles in \cref{eq:amp} by searching for roots of $\det (1-G\, K_\ell)$ in the complex energy plane for every Riemann sheet. 
 Care must be taken in the analytic continuation of the amplitude, especially in coupled channel parameterizations, to identify the relevant Riemann sheets and their proximity to the real axis~\cite{Badalian:1981xj}.

No bound state poles are found in the energy region of interest in any PW for the 1C or 2C best fit results. Attempts to guide these parameterizations to a resonant minimum were done by first fitting only the GlueX total cross section as in~\cite{Strakovsky:2023kqu}. Resulting fits featuring resonant poles were then used as the initial values to fits using the full dataset. Such fits were found to always produce either the quoted nonresonant best fit, or a significantly worse local minimum.

The 3C-NR result presents $S$-wave poles located on Riemann sheets far away from the physical axis, and thus unlikely associated with pentaquark candidates.
On the other hand, the remaining 3C-R fit is found to contain three narrow $S$-wave poles, one of which is compatible with a pentaquark state. We stress that the uncertainties for the 3C fits are the largest and the precise locations of these poles cannot be determined when a detailed bootstrap analysis is performed. Because of this, we report pole positions for the best fit found without uncertainties and focus on the qualitative implications for the $\gamma p \to \jpsi\,p$ reaction.

Using the parameters for the 3C-R best fit, the most relevant pole has a mass of $M = 4211\mev$ and a width of $\Gamma = 48\mev$, placing it between the $\bar{D}\Lambda_c$ and $\bar{D}^*\Lambda_c$ thresholds. In a common notation denoting the relative position to each of the three thresholds~\cite{Badalian:1981xj}, this pole is located on Riemann sheet ($--+$) which is the closest to the physical region. The pole position corresponds to a beam energy of $E_\gamma \sim 9\gev$ and coincides with the structure visible in \cref{fig:integrated_bs}. Besides this, two other poles are found on more remote sheets: one is a mirror pole of the above located on ($-++$), with the same mass and width, and the other pole is located close to the real axis with $M = 4070\mev$ on ($+-+$). 

Again, with the present statistics, definitive conclusions regarding these poles cannot be drawn. Still, these results suggest the recent experimental measurements leave room for the appearance of poles in scenarios with strongly coupled higher channels. Specifically, it suggests the precise lineshape in the dip region is diagnostic of the presence of pentaquarks whose identification may require a sophisticated analysis.

Apart from exciting implications for spectroscopy, the existence of bound states in this region would constitute an unambiguous violation of factorization. 

\section{Summary and outlook}   
\label{S:summary}

In this work we analysed the recent JLab $\jpsi$ photoproduction data near threshold using generic low-energy parameterizations. Most of the literature regarding this reaction relies on specific underlying dynamics of the $\jpsi$-proton interaction, such as the factorization of the nucleon matrix elements. Since these works relate to fundamental properties of the proton, it is important to test the validity of these assumptions against available data. 
We have demonstrated that both integrated and differential cross sections can be described with a small number of partial waves, parameterized with customary low-energy expansions. We have incorporated the effects of nearby open-charm thresholds and extracted quantities which characterize the physics underlying the data. Our results highlight the wide array of physics that may be at play in this energy region.

We have presented four models of increasing complexity, which describe the current data with similar quality but represent different dynamical pictures. We have shown how to extract the elastic $\jpsi\,p$ amplitude from our models while respecting $S$-matrix constraints.
Our analysis indicates that present statistics do not exclude severe violations of factorization and of the Vector Meson Dominance which are usually assumed in the literature.
This may
affect the extraction of the elastic scattering length, total charmonium-nucleon cross section, and proton structure observables, as well as pentaquark searches. 

It is thus crucially important to constrain model parameters with further measurements in order to disentangle the possible physics scenarios and their implications.  In addition to higher statistics, especially to resolve the lineshape around the $9\,\gev$ dip, the measurement of open-charm photoproduction is needed to assess the role of coupled channels. A simultaneous analysis of the $\gamma p \to \jpsi\,p, \bar{D}^{(*)}\Lambda_c$ cross sections would provide a stringent constraint on coupled channel dynamics. Based on the best fit parameters extracted here, we expect a large open-charm cross section $\gtrsim 10\nb$.
Furthermore, studies of photoproduction off nuclear targets may give further constraints on the total \jpsi-nucleon cross section~\cite{HallC:2006}.

Polarization observables were previously proposed as an alternative means to search for pentaquarks~\cite{Winney:2019edt}. This takes advantage of photoproduction facilities' unique capabilities for polarized beam-target setups in accessing helicity dependence. Measuring the $\jpsi$ spin density matrix or spin asymmetries would give access to helicity couplings and may help  further separate the mechanisms at play. Although not considered here, the formalism of \cref{S:formalism} is readily extendable to allow an analysis that includes spin degrees of freedom when such data become available. Our framework can also be applied to the analysis of other vector mesons, such as thfe analysis of $\Upsilon$ photoproduction data when it becomes available in the future. 

The future of heavy meson photoproduction looks promising, with proposals for both upgrading existing experiments, including  measurements in every Hall of JLab~\cite{claspentaquark,HallC:2006,SBS:2018,JeffersonLabSoLID:2022iod}, to further study the near-threshold region, as well as new electron-hadron facilities~\cite{AbdulKhalek:2021gbh,Anderle:2021wcy}. In addition, the proposed $24\gev$ CEBAF upgrade~\cite{Arrington:2021alx,Accardi:2023chb} aims to extend the Jefferson Lab physics program to the charmonium sector. This will give the possibility of also studying higher charmonia in photoproduction~\cite{Winney:2022tky,Albaladejo:2020tzt}, which may give key insight into the role of coupled channels and probe quarkonium wave function dependence to further understand the applicability of VMD in charmonium sectors.  

All codes necessary to reproduce the results of this article are publicly available at~\cite{zenodo}.

\begin{acknowledgments}
We thank S.~Dobbs, K.~Mizutani, and L.~Pentchev for their comments and insight into the GlueX data.
This work was supported by the U.S. Department of Energy contract DE-AC05-06OR23177, under which Jefferson Science Associates, LLC operates Jefferson Lab, 
and also by the U.S. Department of Energy Grant
Nos.~DE-FG02-87ER40365 and DE-FG02-92ER40735,
by U.S. National Science Foundation Grant No.~PHY-2209183,
by the Spanish Ministerio de Ciencia e Innovaci\'on (MICINN) 
Grant Nos.~PID2020-112777GBI00, PID2019–106080GB-C21, and PID2020-118758GB-I00.
DW is supported by National Natural Science Foundation of China Grant No.~12035007 and the NSFC and the Deutsche Forschungsgemeinschaft (DFG, German Research Foundation) through the funds provided to the Sino-German Collaborative Research Center TRR110 ``Symmetries and the Emergence of Structure in QCD'' (NSFC Grant No.~12070131001, DFG ProjectID 196253076-TRR 110).
CFR is supported by Spanish Ministerio de Educaci\'on y Formaci\'on Profesional under Grant No.~BG20/00133.
ANHB is supported by the DFG through the Research Unit FOR 2926 (project number 409651613).
MA is supported by Generalitat Valenciana under Grant No.~CIDEGENT/2020/002.
NH is supported by a Polish research project with
no.~2018/29/B/ST2/02576 (National Science Center). VM is a Serra H\'unter fellow. 
VS acknowledges the support of the USDOE ExoHad Topical Collaboration, contract DE-SC0023598.
This research was supported in part by Lilly Endowment, Inc., through its support for the Indiana University Pervasive Technology Institute.
We acknowledge the computational resources and assistance provided by the Centro de Computaci\'on de Alto Rendimiento CCAR-UNED.
This work contributes to the aims of the U.S. Department of Energy ExoHad Topical Collaboration, contract DE-SC0023598.

\end{acknowledgments}

\appendix
\section{Best fit parameters and uncertainty estimation}
\label{app:errors}
    \begin{table*}
        \centering
        \caption{Best fit results from $\chi^2$ minimization for the four fit results considered. For the 3C model we provide two fits of similar quality and different underlying dynamics. They are labeled 3C-(N)R for (non-)resonant as described in the main text. All numbers are expressed in appropriate GeV units.
        }
        \label{tab:fitvals}
        \begin{ruledtabular}
            \begin{tabular}{l c c cc}
                & \textbf{1C} & \textbf{2C} & \textbf{3C-NR} & \textbf{3C-R}  \\
                \hline
                \# parameters   & $9$      & $13$      & $15$       & $15$      \\
                $\chi^2$     & $166$  & $144$  & $141$   & $143$  \\
                $\chi^2$/dof & $1.25$    & $1.12$    & $1.11$     & $1.13$    \\
                \hline
                $n^{\psi p}_S$ & 0.063  & $0.101$  & $0.105$      & $8.77 \E{-3}$  \\
                $n^{\bar{D}\Lambda_c}_S$ & $-$       & $-$    & $-0.103$  & $9.80$ \\
                $n^{\bar{D}^*\Lambda_c}_S$ & $-$       & $3.214$       & $-0.089$     & $-0.012$  \\
                $\alpha^{\psi p,\psi p}_S$   & $-418.24$ & $-219.68$ & $-258.12$  & $-86.75$  \\
                $\alpha^{\psi p,\bar{D}\Lambda_c}_S$   & $-$       & $-$       & $168.24$   & $-1.34$   \\
                $\alpha^{\psi p,\bar{D}^*\Lambda_c}_S$   & $-$       & $5.00$    & $-132.60$  & $-88.97$  \\
                $\alpha^{\bar{D}\Lambda_c,\bar{D}\Lambda_c}_S$   & $-$       & $-$       & $-135.60$  & $224.25$  \\
                $\alpha^{\bar{D}\Lambda_c,\bar{D}^*\Lambda_c}_S$   & $-$       & $-$       & $235.48$   & $0.081$   \\
                $\alpha^{\bar{D}^*\Lambda_c,\bar{D}^*\Lambda_c}_S$   & $-$       & $47.10$   & $93.98$    & $-294.93$ \\
                $\beta^{\psi p,\psi p}_S$  & $320.76$  & $-180.31$ & $-$        & $-$       \\
                $\beta^{\bar{D}^*\Lambda_c,\bar{D}^*\Lambda_c}_S$   & $-$       & $-145.68$ & $-$        & $-$       \\
                \hline
                $n_P$ & $18.3\E{-3}$  & $14.6\E{-3}$ & $16.1\E{-3}$   & $14.02\E{-3}$   \\
                $\alpha_P$  & $-133.77$ & $-44.00$ & $-61.24$   & $-87.80$  \\
                \hline
                $n_D$   & $3.08\E{-3}$  & $3.03\E{-3}$ & $3.63\E{-3}$     & $3.65\E{-3}$    \\
                $\alpha_D$  & $-36.32$  & $-2.34$  & $-4.77$    & $-16.55$  \\
                \hline
                $n_F$ & $0.81\E{-3}$  & $0.69\E{-3}$ & $0.52\E{-3}$   & $0.66\E{-3}$   \\
                $\alpha_F$  & $-25.91$  & $-6.01$  & $3.14$     & $-10.17$ 
            \end{tabular}
        \end{ruledtabular}
    \end{table*}
\begin{figure}[t]
    \centering
    \includegraphics[width=\columnwidth]{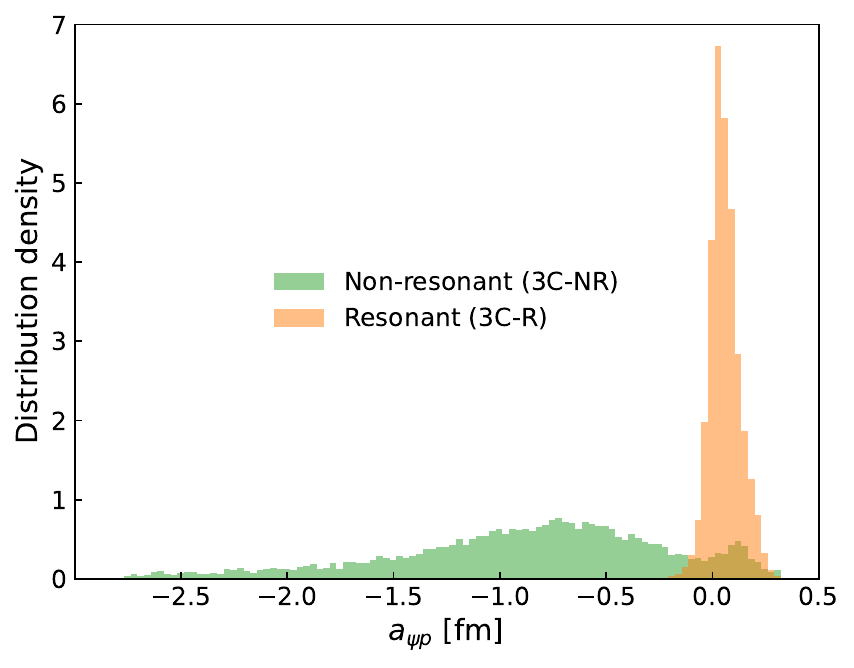}
    \caption{Distribution of the $a_{\psi p}$ values obtained from the bootstrap and used to compute their uncertainties in \cref{tab:results} for the 3C-R and 3C-NR. The area of each distribution is normalized to unity. A secondary peak is clearly visible near $a_{\psi p} \simeq 0.1 \fm$ in the 3C-NR distribution, and is likely due to cross-contamination from the other solution.}
    \label{fig:a3chistogram}
\end{figure}

In \cref{tab:fitvals} we provide the best fit results from the $\chi^2$ minimization of the four models considered.
These parameters are highly correlated and, because their absolute size is not of interest \textit{per se}, we do not show uncertainties. Instead, for each fit case, we compute 68\% CL uncertainties for the curves of both integrated and differential cross sections and 90\% CL for the extracted dynamical quantities reported in \cref{tab:results}. Using a bootstrap analysis~\cite{JPAC:2021rxu}, all sources of experimental uncertainties, \ie statistical, uncorrelated systematics, and correlated systematics, are propagated to each quantity of interest. To compute the confidence intervals we perform $10^{4}$ bootstrap fits to obtain the distribution for each quantity. 

Because this minimization is ill-posed, it is possible for some bootstrap fits to end in local minima which are quite far away from the best fit. Such outliers are clearly separated from the rest of the distribution and highly affect the extracted mean and standard deviation unrealistically. 
In order to handle this, for the error estimations in \cref{tab:results}, we use an iterative process to prune outliers and achieve a more realistic error estimation. For a given distribution we compute the mean and standard deviation and remove any values $4\sigma$ away from the mean. This step is repeated until all values in the remaining distribution lie within $4\sigma$. In all cases, this pruning procedure removes at most $7.5\%$ of the initial $10^4$ bootstrap fits before convergence is achieved. With the final pruned distribution, the 90\% CL interval is computed as the range between the upper and lower 5\% tails.

The 3C parameterization has to be considered with care due to the presence of the two quoted minima (\ie the 3C-NR and 3C-R). During the bootstrap calculation, cross-contamination between the two solutions is possible, as can be see in \eg the histograms in \cref{fig:a3chistogram}. Similar distribution shapes are also seen in the $R_\text{VMD}$ values for these cases. Nevertheless the two solutions can be clearly separated at a 68\% CL and the overlap remains relatively mild when considering the 90\% CL, thus we expect the uncertainty estimation to be reliable. 

\bibliographystyle{apsrev4-1.bst} 
\bibliography{5q}

\end{document}